\RequirePackage[T1]{fontenc}
\documentclass[12pt]{article}
\pdfoutput=1 % if your are submitting a pdflatex (i.e. if you have images in pdf, png or jpg format)
\usepackage[height=8.85in,width=6.45in]{geometry}

\usepackage[utf8]{inputenc}
\usepackage{amsmath}
\usepackage{amssymb}
\usepackage{mathtools}
\numberwithin{equation}{section}
\usepackage{slashed}
\usepackage{braket}
\usepackage[svgnames]{xcolor}
\usepackage[colorlinks,linktocpage=true,citecolor=DarkGreen,linkcolor=FireBrick]{hyperref}
\usepackage{cite}
\usepackage{graphicx}
\usepackage{tikz}
\usepackage{tikz-cd}
\usepackage{times}
\usepackage[scaled]{couriers}
\usepackage{bm}
\usepackage{subfig}
\usepackage{mathrsfs}
\usepackage{amsthm}

\usepackage{xcolor}
\usepackage{mdframed}

\usepackage{amsmath}
\usepackage{amssymb}
\usepackage{amsfonts}
\usepackage{graphicx}
\usepackage{xcolor}
\usepackage{xfrac}
\usepackage{comment}
\usepackage{pifont}
\usepackage{physics}
\usepackage{fourier}
\usepackage{hyperref}
\usepackage{bm}
\usepackage{enumitem}

\definecolor{rossoferrari}{HTML}{D9073D}
\definecolor{mediumblue}{HTML}{0000CD}
\definecolor{forestgreen}{HTML}{228B22}
\definecolor{desy_blue}{HTML}{009EE2}
\definecolor{desy_orange}{HTML}{FD8800}
\definecolor{light_pink}{rgb}{1,0.4,0.4}
\definecolor{light_blue}{rgb}{0.284602,0.317763,0.963947}
\hypersetup{setpagesize=false,bookmarksnumbered=true,bookmarksopen=true,%
colorlinks=true,linkcolor=light_blue,urlcolor=rossoferrari,citecolor=rossoferrari,linktocpage=false}

\newcommand{\gsim}{ \mathop{}_{\textstyle \sim}^{\textstyle >} }

\usepackage{slashed}

\def\cO{{\cal O}}

\def\bZ{{\mathbb Z}}

\def\sc{{\mathsf c}}

\def\SU{\mathrm{SU}}
\def\O{\mathrm{O}}
\def\SO{\mathrm{SO}}
\def\Sp{\mathrm{Sp}}

\def\Spin{\mathrm{Spin}}

\def\beq#1\eeq{\begin{align}#1\end{align}}

\newcommand{\eV}{ \ {\rm eV} }

\newcommand{\GeV}{\  {\rm GeV} }

\newcommand{\lmk}{\left(}  
\newcommand{\rmk}{\right)}

\newcommand{\Mpl}{M_{\rm Pl}}

\newcommand{\eq}[1]{Eq.~(\ref{#1})}

\begin{document}

\begin{titlepage}

\begin{flushright}
TU-1203
\end{flushright}

\vskip 3cm

\begin{center}

{\Large \bfseries Dark baryon from pure Yang-Mills theory
and its GW signature from cosmic strings}

\vskip 1cm

Masaki~Yamada$^{1,2}$
and 
Kazuya~Yonekura$^{1}$

\vskip 1cm

\begin{tabular}{ll}
$^1$ Department of Physics, Tohoku University, Sendai 980-8578, Japan
\\
$^2$ FRIS, Tohoku University, Sendai, Miyagi 980-8578, Japan
\end{tabular}

\vskip 1cm

\end{center}

\noindent
We point out that $\SO(2N)$ pure Yang-Mills theory provides a candidate for dark matter (DM) without the explicit need to impose any additional symmetry. The DM candidate is a particular type of glueball, which we refer to as a baryonic glueball, that is naturally stable and produced by a novel production mechanism for a moderately large $N$. In this case, the intercommutation probability of cosmic strings (or macroscopic color flux tubes) is quite low, which offers characteristic gravitational wave signals to test our model. In particular, our model can simultaneously account for both abundance of DM and the recently reported gravitational wave signals detected in pulsar timing array experiments, including NANOGrav. 
\end{titlepage}

\setcounter{tocdepth}{2}

\newpage

\tableofcontents

\section{Introduction}
History has shown that physics has developed through the pursuit of simplicity and naturalness, and one would hope that the mysteries of particle physics and cosmology will also be solved in this direction. 
The pure Yang-Mills (YM) theory embodies these attributes 
and we consider the gauge group SO($2N$) in this paper. 
Despite its conceptual simplicity, this theory offers a rich and diverse phenomenology in both particle physics and cosmology. 
The gauge coupling is asymptotically free and the theory is confined at a low energy scale. 
The model predicts the formation of macroscopic color flux tubes and various glueballs after the confinement phase transition.

Gauge theories have color flux tubes as dynamical objects, and they can be stable cosmic strings depending on the gauge group $G$ and the matter content. This was originally pointed out by Witten \cite{Witten:1985fp} and examined in much more detail from a modern viewpoint in Refs.~\cite{Yamada:2022imq,Yamada:2022aax}. 
In accordance with the theory of holographic duality, 
they can also be interpreted as cosmic superstrings on the graivty side, 
which supports the notion that the cosmic strings have a small intercommutation probability. 
This leads to the prospect of observing gravitational wave (GW) signals originating from the dynamics of cosmic strings as suggested in Refs.~\cite{Yamada:2022imq,Yamada:2022aax}.%
\footnote{
GWs from the dynamics of the phase transition have been considered in Refs.~\cite{Reichert:2021cvs,Morgante:2022zvc,He:2022amv,Reichert:2022naa}. However, the resulting signals do not fall within the frequency range of interest in our case. 
}

Glueballs are color-singlet bound states consisting of gluons and play an intriguing role in our model. 
The lightest glueball tends to dominate the energy density of the Universe shortly after the phase transition~\cite{Morningstar:1999rf,Lucini:2010nv,Curtin:2022tou}. 
However, these glueballs are unstable via higher-dimensional operators, leading eventual reheat of the Universe via decay into Standard Model (SM) particles~\cite{Juknevich:2009ji,Juknevich:2009gg,Halverson:2016nfq,Asadi:2022vkc}.%
\footnote{
See Refs.~\cite{Faraggi:2000pv,Feng:2011ik,Boddy:2014yra,Boddy:2014qxa,Soni:2016gzf,Kribs:2016cew,Forestell:2016qhc,Soni:2017nlm,Forestell:2017wov,Jo:2020ggs,Carenza:2022pjd,Carenza:2023shd} 
for a discussion on long-lived glueball DM, where they considered a dynamical scale that is significantly smaller than ours. 
}
In contrast, the $\SO(2N)$ pure YM theory contains another, qualitatively different type of glueballs, which we refer to as baryonic glueballs. 
The lightest baryonic glueball is extremely stable because they are protected by an accidental $\bZ_2$ symmetry. The dimensions of operators that explicitly break $\bZ_2$ are at least $2N$ which is quite high for moderately large $N$.
Therefore, 
the baryonic glueball is a promising candidate for dark matter (DM). 
We propose that baryonic glueballs for large $N$ 
may be produced at the confinement phase transition via a variant of the Kibble-Zurek mechanism.%
\footnote{
Hidden monopoles, generated by a similar mechanism, are considered as a DM dandidate in Refs.~\cite{Murayama:2009nj,Baek:2013dwa,Khoze:2014woa,Kawasaki:2015lpf}.
}

In this paper, we determine a parameter space in which the baryonic glueball can account for the abundance of DM and predict GW signals from cosmic strings. 
Our model is characterized by its simplicity and is defined by merely two free parameters, $N$, and the dynamical (or confinement) scale $\Lambda$. 
We find that, subject to some $\mathcal{O}(1)$ unknown theoretical factors, 
the recentely reported signals of GWs by 
the North American Nanohertz Observatory for Gravitational Waves (NANOGrav)~\cite{NANOGrav:2023gor}, 
Parkes Pulsar Timing Array (PPTA)~\cite{Reardon:2023gzh}, 
European PTA (EPTA)~\cite{Antoniadis:2023ott}, 
and Chinese PTA (CPTA)~\cite{Xu:2023wog} 
can be collectively explained in our model.

The rest of this paper is organized as follows. In Sec.~\ref{sec:theory}, we explain field theoretical aspects of pure SO($2N$) YM theory and disucss the longevity and formation mechanism of baryonic glueball. In Sec.~\ref{sec:glueball}, we consider the thermal history of the model. The unstable glueballs dominantly form at the confinement phase transition but eventually decay into the SM particles at a low temperature. We also estimate the abundance of baryonic glueballs and show that it can explain the DM. In Sec.~\ref{sec:cosmicstring}, we calculate the GW spectrum that is emitted from cosmic string loops and compare it with the existence data and future sensitivities for GW experiments. Sec.~\ref{sec:discussion} comprises the discussion and conclusions of the paper.

\section{Field theoretical aspects of pure Yang-Mills theory}
\label{sec:theory}

\subsection{Stability and thermal relic of baryonic glueballs} 

We first explain the theoretical foundation of glueballs in the $\SO(2N)$ YM theory. We denote the gauge field strength of $\SO(2N)$ as $(F_{\mu\nu})_{ij}$, where $\mu,\nu,\cdots$ represent spacetime indices, and $i,j,\cdots$ denote gauge indices ranging from $1$ to $2N$. 
$(F_{\mu\nu})_{ij}$ is antisymmetric with respect to $i \leftrightarrow j$. The lightest glueball may be created by the operator $O=(F_{\mu\nu})_{ij} (F^{\mu\nu})_{ij} $. This glueball can decay into SM particles through the interaction of a higher dimensional operator, such as $\frac{1}{M^2} |h|^2 O$, where $h$ denotes the SM Higgs field and $M^{-2}$ represents the coupling. 

However, another type of glueballs exists, which we refer to as ``baryonic glueballs''. We can construct gauge invariant operators by using the totally antisymmetric tensor $\epsilon_{i_1   \cdots  i_{2N}} $ of $\SO(2N)$ as
\beq
B_{\mu_1  \cdots  \mu_{2N}} = 
\epsilon_{i_1   \cdots  i_{2N}} (F_{\mu_1\mu_2})_{i_1i_2} \cdots (F_{\mu_{2N-1}\mu_{2N}})_{i_{2N-1}i_{2N}} .\label{eq:baryonic}
\eeq
These operators, when acted on the vacuum state $|\Omega\rangle$, create baryonic glueball states.

There are two reasons that baryonic glueballs are phenomenologically interesting for a large $N$. 
First, they are stable because of an accidental symmetry. We can consider not just $\SO(2N)$ but also $\O(2N)$ transformations. We have taken the gauge group to be $\SO(2N)$. Then the quotient
\beq
\O(2N)/\SO(2N) =\bZ_2,
\eeq
(i.e. $\O(2N)$ transformations modulo $\SO(2N)$ transformations) is a global symmetry.
Baryonic glueballs are charged under this $\bZ_2$ global symmetry because 
$\epsilon_{i_1  \cdots  i_{2N}}$ is not invariant; instead, it changes its sign under the nontrivial element of $\O(2N)/\SO(2N)$. Here, $\bZ_2$ is just an accidental symmetry, and we do not impose it by hand. It can be explicitly broken only by very higher dimensional operators like \eqref{eq:baryonic} whose dimensions are at least $2N$.%
\footnote{
After submitting our work, we noticed a paper that had discussed the longevity of a glueball in SO(2N) theory~\cite{Gross:2020zam}, where they considered different production mechanisms.
}
Hence, if $N$ is large, the accidental $\bZ_2$ symmetry is a very good approximate symmetry and the lightest baryonic glueball is naturally stable, without imposing any global symmetry by hand.

The second reason for the interest in baryonic glueballs 
is owing to their unique production characteristics. 
In general $\SU(N)$ or $\SO(N)$ gauge theories, it is known that baryon masses are of order $N$ in the large $N$ limit~\cite{Witten:1979kh,Witten:1998xy}. Including an ${\cal O}(N^0)$ term, the lightest baryon mass can be parametrized as $m_B \sim (N-a)\Lambda$, where $\Lambda$ is the typical mass scale of the gauge theory under consideration and $a$ represents an order one constant. Moreover, let $T_c$ denote the critical temperature of the phase transition from deconfinement to the confinement phase. 
In the large $N$ limit, 
$T_c$ is of order $N^0$ and may be parametrized as $T_c \sim b^{-1} \Lambda$, where $b$ is a dimensionless constant. 
Baryon particles exist only in the confinement phase, with the maximum temperature of the confinement phase being $T_c$. Consequently, the Boltzmann suppression factor is proportional to
$\exp( - m_B/T_c)$ which is exponentially small in the large $N$ limit such as 
\beq
\exp( - m_B/T_c) \sim \exp\left( -b(N-a) \right) \label{eq:Botzmann}
\eeq
where $b = \Lambda/T_c$.% and we have neglected $\exp(ab)$. 

Due to strong dynamics, it is hard to calculate the precise value of the dimensionless constant $b$. A very crude argument might be given as follows. Finite temperature field theories can be described by an $S^1$ compactification of Euclidean time with a period $1/T$, where $T $ is the temperature. The Kaluza-Klein mass scale of the $S^1$ compactification is $2\pi T$. The phase transition might occur when this mass scale is comparable to the hadron mass scale $\Lambda$. Consequently, $2\pi T_c \sim \Lambda$ or $b \sim 2\pi$ up to an ${\cal O}(1)$ factor. There is also another crude argument. By quantizing an F-string, we get a Regge trajectory of the form $m_n^2 \sim \alpha'^{-1} n +{\rm const.}$, where $m_n^2$ is the mass squared of a hadron labeled (among other things) by an integer $n$, and $\alpha'$ is the Regge slope which determines the mass scale of hadrons. The Hagedorn temperature $T_H$ is related to $\alpha'$ by $2\pi T_H =  {\cal O}(1)\alpha'^{-1/2}$, where the ${\cal O}(1)$ factor depends on the type of the F-string (i.e. its worldsheet theory). By identifying $\Lambda \sim \alpha'^{-1/2}$ and $T_c \sim T_H$, we again get $2\pi T_c \sim \Lambda$ up to an ${\cal O}(1)$ factor. Actually, $ T_c<T_H$ for a first order transition~\cite{Aharony:2003sx}, which implies $b \gsim 2\pi $.

The value $b \sim \Lambda/T_c \sim 2\pi$ seems to be reasonable in the case of $\SU(N)$ gauge theories with $N_f$ flavors of quarks. First, note that when $N=2$ (i.e. $\SU(2)=\Sp(1)$ gauge theory), the lightest baryon is merely a part of the massless Goldstone bosons of chiral symmetry breaking $\SU(2N_f) \to \Sp(N_f)$. Thus, it is reasonable to take $m_B \sim (N-2)\Lambda$.
For $N=3$, the proton mass in the real QCD is approximately $938~{\rm MeV}$, which we might use as an estimate of $m_B\sim (3-2)\Lambda= \Lambda$. This value is indeed comparable to non-Goldstone meson masses and the Regge slope. The critical temperature $T_c$ for the real QCD has been estimated by lattice simulations to be around $T_c \sim 160~{\rm MeV}$~\cite{Bhattacharya:2014ara}. Thus we might estimate $b= \Lambda/T_c \sim 938/160 \sim 6 $, which is close to $2\pi$. All these arguments are very crude, but the point is that $b=\Lambda/T_c$ may be more appropriately regarded as $2\pi \cdot {\cal O}(1)$ and may be a bit large. Then the Boltzmann factor $\exp(-bN)$ may be extremely small for a moderately large $N$ of the order ${\cal O}(10)$. Therefore, we assume in the following that $N$ is large enough so that the thermal production of baryonic glueballs in our $\SO(2N)$ gauge theory is negligible.

\subsection{Production of baryonic glueballs via variant of Kibble-Zurek mechanism}

\begin{figure}
    \centering
            \includegraphics[width=1.0\hsize]{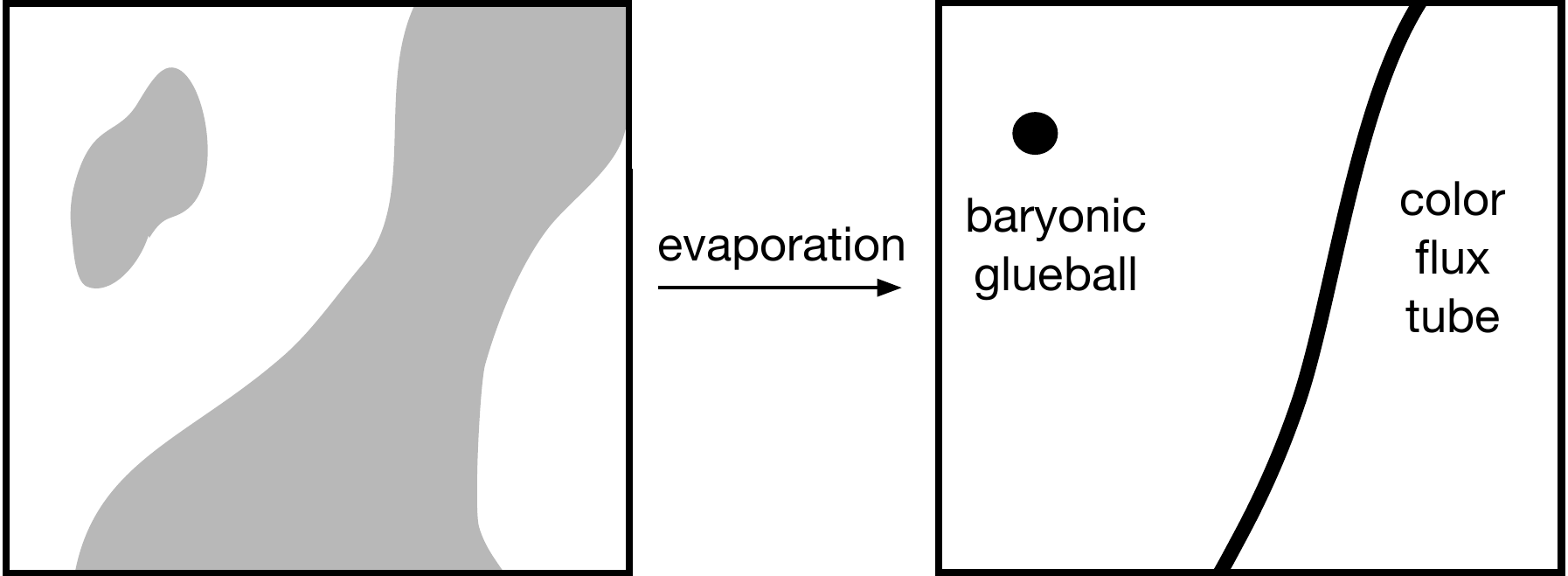}
    \caption{
In the left part of the figure, gray regions represent the deconfined phase while white (uncolored) regions represent the confined phase. The deconfined regions form dynamical objects bound by strong interactions. They evaporate by emitting ordinary glueballs, and by some probabilities they leave baryonic glueballs or color flux tubes.
}
    \label{fig:deconf}
\end{figure}

While baryonic glueballs are not thermally produced, they may still be generated during the phase transition from deconfinement to confinement phase by a variant of the Kibble-Zurek mechanism as follows. (See also \cite{Witten:1984rs} for a similar, though not identical, situation.) The phase transition proceeds by first creating bubbles of confinement phase inside the deconfined thermal bath. After some time, the confined regions occupy more than half of the volume of the universe, and some of the deconfined regions turn into isolated bubbles. 
In the bulk of the confined regions, there are almost no baryonic glueballs due to the Boltzmann suppression \eqref{eq:Botzmann}. However, there is no such suppression inside the deconfined bubbles, and hence each bubble has $\bZ_2$ charge which is either $0$ or $1$ determined randomly by the number of gluons charged under $\bZ_2$. Each bubble eventually evaporates and it must leave a baryonic glueball if it has charge $1 \in \bZ_2$. See Figure~\ref{fig:deconf}.

More precisely, each isolated deconfined bubble needs to be color-singlet. Thus it is required that gauge indices $i, j, \cdots $ of gluons inside the thermal bath of a deconfined bubble are contracted either with the Kronecker delta $\delta_{ij}$ or the totally antisymmetric tensor $\epsilon_{ij\cdots}$. Let $n_i~(i=1,\cdots,2N)$ be the number of times the index $i$ appears as a color index of gluons, and let $n_{\rm total} = \sum_{i=1}^{2N} n_i$. The total number of gluons inside the thermal bath is $n_{\rm total} /2$ since each gluon has two indices. Gauge invariance implies
\beq
n_1 \equiv n_2 \equiv \cdots \equiv n_{2N} \mod 2.\label{eq:gaugeinvariance}
\eeq
Namely, all of $n_i$ are simultaneously even or odd.
Let $n$ be 
\beq
n \equiv n_i \mod 2 \quad \text{(for any $i$).}
\eeq
This $n $ (which is either 0 or 1) is the $\bZ_2$ charge of the deconfined bubble mentioned above. When the number of glouns is small, $n_{\rm total}  \ll N $, the $\bZ_2$ charge must vanish, $n = 0 $, as one can see from the condition \eqref{eq:gaugeinvariance}. On the other hand, when $n_{\rm total} \gg N$, it is intuitively clear that the two cases $n= 0$ and $n = 1$ appear with almost the same probabilities. This intuition may be confirmed as follows. Let $Z= \tr e^{-\beta H}$ be the partition function of the thermal bath inside the deconfined bubble, and let $Z'= \tr (-1)^n e^{-\beta H}$ where $(-1)^n$ is regarded as the operator of the $\bZ_2$ symmetry acting on the Hilbert space.
Let $p_n~(n=0,1)$ be the probability that the $\bZ_2$ charge is $n$. We have
\beq
p_0 = \frac{Z+Z'}{2Z}, \qquad p_1 = \frac{Z-Z'}{2Z}.
\eeq 
The factor $(-1)^n$ can be regarded as a Wilson line of a nontrivial element of $\O(2N)$ around the thermal circle, and we can compute $Z$ and $Z'$ by a Euclidean path integral as in \cite{Gross:1980br}.\footnote{A detailed exposition of this kind of computations is also given e.g. in Appendices A and B of \cite{Kobayashi:2023ajk}.} Then we find that $\log(Z'/Z) \sim -  N  T^3 V$ where $V$ is the volume of the deconfined bubble which is assumed to be large, and $T=\beta^{-1}$ is the temperature. The factor $N$ comes from the fact that, if we take e.g. $n=n_1$, then there are $2N-1$ species of gluons charged under $(-1)^{n_1}$ whose indices are $(1,j)$ for $j=2,\cdots, 2N$. 
(We are assuming $N \gg 1$ and hence $2N-1 = \cO(1) \cdot N$.) Recalling that the number of gluons is $n_{\rm total} \sim N^2  T^3 V$, we get
\beq
\frac{Z'}{Z} \sim \exp \left( -\cO(1) \cdot \frac{n_{\rm total}}{N} \right).
\eeq
This result is valid if $V$ is large enough so that the thermodynamic computation is applicable.
From this result, we see that $p_0 \simeq p_1 \simeq \frac12$ if $n_{\rm total} \gg N$. This computation is done only at the leading order of the coupling expansion, but we believe that the conclusion is qualitatively valid even in the strongly coupled plasma. 

Deconfined bubbles, or plasma balls~\cite{Aharony:2005bm}, are characteristic to large $N$ confining gauge theories, and they are believed to be metastable when $N$ is large. At the beginning of the formation, deconfined bubbles may have complicated shapes. Due to attractive forces of color fluxes, however, they form bound states and may not easily be torn apart (see also the next paragraph.)
They reach their metastable, spherical configurations by emitting ordinary (rather than baryonic) glueballs. After that, they slowly decay by further emitting glueballs around the critical temperature $T_c$. Notice that the energy density of a deconfined bubble is $\cO(N^2)$ in the large $N$ limit, while glueball emission is $\cO(N^0)$. See \cite{Aharony:2005bm} for more details. 

Holographic interpretation of the above process is as follows. Deconfined bubbles are dual to black holes in the gravity side~\cite{Witten:1998zw,Aharony:2005bm}, and in our case they are surrounded by the thermal bath. Black holes reach their stable shapes by emitting ``gravitational waves'' (in the holographic sense) or gravitons, which are dual to glueballs. In general, black holes are strongly bound and may not easily be torn apart\,\footnote{For instance, when the dual gravity theory is approximated by Einstein gravity, the Hawking area theorem gives a strong restriction to such processes. Pure YM theories are not dual to Einstein gravity, so this point is not completely obvious. } aside from some quantum gravitational effects that are suppressed in the large $N$ limit.\footnote{Recall that the ``Newton constant'' $G_{\rm hol}$ in holographic dual is of order $G_{\rm hol} \sim N^{-2}$. The first order phase transition from the deconfined to confined phase is an example of a quantum gravitational effect and its probability is suppressed as $e^{-S} = e^{-\cO(1) \cdot N^2}$, where $S$ is the bounce action which is of order $S \sim \frac{1}{G_{\rm hol}} \sim N^2$.}
As time passes, black holes evaporate by emitting Hawking radiation and leave either nothing or a D-brane or other brane. Some D-brane is the dual of the baryonic glueball~\cite{Witten:1998xy}.

The mechanism discussed here may also imply the production of cosmic strings, which are argued in a different way by using electric-magnetic duality in Refs.~\cite{Yamada:2022imq,Yamada:2022aax}. For this argument, we consider a tube, rather than a bubble, of deconfinement phase. In the deconfinement phase, there are random color fluxes throughout the thermal bath. Formally we can describe the randomness of color fluxes by saying that the 1-form center symmetry is spontaneously broken in the deconfinement phase~\cite{Gaiotto:2014kfa}. (For a detailed review of the concept of 1-form symmetries in the context of cosmic strings, see \cite{Yamada:2022imq}.) As the region of deconfinement phase becomes smaller, some regions are of the form of tubes, and let us consider a cross section of such a tube.  (Recall that charges of 1-form symmetries are associated to cross sections rather than volumes.) The charge of the 1-form center symmetry associated to this cross section takes random values. This is analogous to the random $\bZ_2$ charge discussed above. If the charge is nonzero, there remains a string after the deconfined tube evaporates. See Figure~\ref{fig:deconf}.
Holographically, we may interpret it as a process that black strings evaporate and leave either nothing or a (F- or D-) string. The argument here gives independent evidence for the claim made in \cite{Yamada:2022imq,Yamada:2022aax} that macroscopic color flux tubes are produced after the phase transition.

We believe that the above scenario of the production of baryonic glueballs and color flux tubes is reasonable due to the properties of large $N$ gauge theories discussed above. Needless to say, strong interactions make it hard to actually prove it, so it would be very interesting to further investigate the validity of the scenario.

\section{Cosmology of glueballs}
\label{sec:glueball}
\subsection{Unstable glueball domination and decay}
Now let us consider the thermal history of the model. 
We assume that the reheating temperature after inflation is higher than the confinemnt scale. As the temperature decreases owing to the redshift, 
the phase transition from the deconfinement phase to the confinement phase occurs at $T = T_c$. (Due to supercooling, the cosmological critical temperature may be smaller than the field theoretical critical temperature, but they are expected to be of the same order.) 
After this phase transition, 
we expect that the majority of the energy density of gluons is converted into that of glueballs (which we emphasize are distinct from bayonic glueballs). 
The energy density of glueballs is estimated as 
\beq
 \rho_g(T_c) \simeq 
 \frac{g_A \pi^2 T_c^4}{30}, 
 \label{eq:rhogluball}
\eeq
where $g_A = 2N(2N-1)$ represents the degrees of freedom of the SO($2N$) gauge boson. 
Since we expect $T_c \sim \Lambda/(2\pi)$ and the glueballs become non-relativistic shortly after formation, they dominate the Universe after the phase transition. 

However, 
if the lightest glueball is thermalized at a temperature around $T \sim T_c$, its energy density should be of order $m_g T_c^3$, where $m_g$ ($\sim \Lambda$) is the lightest glueball mass. 
\eq{eq:rhogluball} exceeds this value for a sufficiently large $N$ and the remaining fraction of the energy density should be stored by some other form. 
The latter energy density is expected to be converted to that of glueballs later. Presumably the energy is contained in the deconfined bubbles discussed above, which constantly emit glueballs.

The lightest glueball is singlet and has self-interactions that may reduce its number density. 
From the standard large $N$ counting, the amplitude for the $3\to2$ scattering scales as $N^{-3}$ and hence the rate is 
\beq
 \Gamma_{3\to2} \sim \frac{n_g^2}{N^{6} \Lambda^{5}}, 
\eeq
where $n_g$ is the number density of glueballs. 
Substituting $n_g = \rho_g / m_g$,
this rate may be smaller than the Hubble parameter at the phase transition for the parameter of our interest.
We can thus neglect this effect and use $\rho_g (T) \propto a^{-3}$ until the glueballs decay.

The glueballs eventually decay into radiation via higher-dimensional operators as discussed above. 
The decay rate is estimated as~\cite{Halverson:2016nfq}
\beq
 \Gamma_d \sim \frac{\Lambda^6}{4 \pi m_g M^4}, 
\eeq
where $m_g$ ($\sim \Lambda$) denotes the lightest gluball mass. 
The decay temperature of the glueball, $T_d$, can be estimated by $\Gamma_d = H(T_d)$, where $H$ is the Hubble parameter. It is calculated as 
\beq
 T_d &\simeq \lmk \frac{90}{g_* \pi^2} \rmk^{1/4}
 \sqrt{\frac{\Mpl}{4\pi m_g}} \frac{\Lambda^3}{M^2} 
 \\
 &
 \simeq 1.3 \times 10^4 \GeV \lmk \frac{\Lambda}{10^{13} \GeV} \rmk^{5/2} 
 \lmk \frac{M}{\Mpl} \rmk^{-2}
 \lmk \frac{m_g}{\Lambda} \rmk^{-1/2}, 
\eeq
where $g_*$ ($\simeq 106.75$) is the effective degrees of freedom for relativistic particles at the decay temperature. 
The decay of glueballs generates entropy and dilutes cosmological relics, including baryonic glueballs.

\subsection{Production of baryonic glueballs}
We have assumed that the confinement phase transition is of the first-order, and it proceeds via the nucleation of bubbles for the confiement phase. 
Around $T \sim T_c$, 
the nucleation rate is symbolically represented as 
\beq
 \Gamma_{\rm PT} \sim \Lambda^4 e^{-S(T)}, 
\eeq
where $S(T)$ represents the exponent of the nucleation rate. 
Although we cannot compute the nucleation rate directly, we ultilized this representation to demonstrate the dynamics of the cosmological phase transition.%
\footnote{
The exponent for the nucelation rate was estimated in Refs.~\cite{Bigazzi:2020phm,Bigazzi:2020avc,Halverson:2020xpg,Huang:2020crf,Reichert:2021cvs,Morgante:2022zvc,He:2022amv,Reichert:2022naa}. 
However, these calculations involve non-trivial assumptions that lack comprehensive theoretical justification. 
}
Still, we expect $S(T) \approx N^2 \hat{S}(T)$ with $\hat{S}$ being almost independent of $N$ by the general large $N$ counting.

The critical temperature for the phase transition $T_c$ is determined by $\Gamma_{\rm PT} \sim H^{4} (T_c)$. 
Since $T_c \sim \Lambda/(2\pi)$ in our case, we obtain $S(T_c) \simeq 4 \ln \Lambda / H(T_c) \sim 60$ for $\Lambda \sim 10^{13} \GeV$. 
The duration of the phase transition $\beta^{-1}$ can be estimated using the time derivative of the nucleation rate such as 
\beq
 \beta = \frac{d \ln \Gamma_{\rm PT}}{d t} \simeq n S(T_c) H_{\rm PT}, 
\eeq
where we assume a power low dependence for $S(T) \propto T^{n}$ with some constant $n$. 
We note that $\hat{S}(T_c) \approx S(T_c)/N^2 \sim \mathcal{O}(1)$ for $N = \mathcal{O}(10)$, which implies a regular behavior on $T$ dependence of $\hat{S}(T)$ at $T \simeq T_c$.
We thus expect $n = \mathcal{O}(1)$
and accordingly $\beta = (10\,-\,100) H_{\rm PT}$. 
The phase transition can be completed within the timescale of $\beta^{-1}$, implying that 
the typical size of bubbles is of order $\beta^{-1}$ at the end of phase transition. 
The Hubble parameter at the critical temperature, 
$H_{\rm PT}$, is determined by radiation in both SM and dark sectors and 
depends on the temperature differences in those sectors. Hereafter we neglect SM contribution and use $H_{\rm PT}^2 = \rho_g(T_c) / (3 \Mpl^2)$ for simplicity.

As discussed above, 
we can estimate the number density of baryonic glueballs at the end of the phase transition by the variant of the Kibble-Zurek mechanism such as 
\beq
n_b \simeq c \beta^{3}, 
\eeq
where $c$ is a numerical constant 
and 
we expect $\mathcal{O}(0.01\,\text{-}\,1)$. 
The energy density $\rho_b$ is given by $m_B n_b$, 
where $m_B$ ($\sim N \Lambda$) denotes the baryonic glueball mass. 
The energy density of baryonic glueballs decreases as $a^{-3}$, where $a$ is the scale factor.

After the decay of glueballs, 
the abundance of baryonic glueball is given by 
\beq
 \frac{\rho_b}{s} &\simeq \frac{g_* \pi^2 T_d^4/30}{2 \pi^2 g_{*s} T_d^3/45} 
 \frac{\rho_b}{\rho_g} 
 \\
 &\simeq 0.4 \eV \times N^{2} \lmk \frac{c}{0.1} \rmk \lmk \frac{\beta}{10^2 H_{\rm PT}} \rmk^3
 \lmk \frac{\Lambda}{10^{13} \GeV} \rmk^{11/2}  
 \nonumber\\
 &\qquad \times \lmk \frac{M}{\Mpl} \rmk^{-2} 
 \lmk \frac{m_g}{\Lambda} \rmk^{-1/2}
 \lmk \frac{m_B}{N \Lambda} \rmk 
 \lmk \frac{T_c}{\Lambda/(2\pi)} \rmk^{2}, 
\eeq
where $g_{*s} \simeq g_*$ and we assume $N \gg 1$. 
The observed DM abundance of $\rho_{\rm DM} /s \simeq 0.4 \eV$~\cite{Planck:2018vyg} can be accounted for by the baryonic glueball when $\Lambda \sim 10^{12}\,\text{-}\, 10^{14} \GeV$ for $N \sim 10$ within theoretical uncertainties regarding several $\mathcal{O}(1)$ factors.

\section{Dynamics of cosmic strings}
\label{sec:cosmicstring}

\subsection{Properties of cosmic strings}
The pure YM theory based on the $\SO(2N)$ gauge group has two types of stable color flux tubes that we may refer to as ``F-string'' and ``D-string''~\cite{Yamada:2022imq,Yamada:2022aax}. The F-string (resp. D-string) is a color flux tube created by an external charge in the fundamental (resp. spinor) representation of $\SO(2N)$.
If we introduce dynamical matter fields in the fundamental (resp. spinor) representations, then the F-string (resp. D-string) can end on these matter particles and become unstable. However, in the pure YM theory, these strings cannot end on anything and hence they are stable. See \cite{Witten:1985fp,Yamada:2022imq,Yamada:2022aax} for more detailed discussions. It is also possible to consider a gauge theory in which only matter fields in the fundamental representation are introduced, in which case the D-string remains stable.

The F- and D-strings exhibit distinct properties compared to Abelian-Higgs cosmic strings, especially a small intercommutation probability. 
Their string tensions and intercommutation probabilities are summarized as 
\beq
\label{eq:muP}
\begin{array}{c|c|c}
&\text{F-string}& \text{D-string} \\ \hline
\mu~~ &\Lambda^2& N\Lambda^2  \\ \hline
P~~ &N^{-2} & \exp(-c N)  %\\ \hline
\end{array}
\eeq
within $\mathcal{O}(1)$ theoretical uncertainty 
in the large-$N$ limit, at least for $\mathcal{O}(1)$ relative velocity. 
Although the intercommutation probability of D-strings has various theoretical uncertainty, 
it may be reasonable to assume $P \sim 10^{-3}\,\text{-}\,10^{-1}$ for $N \sim 10$. 
If we parametrize $\mu = c_\mu \Lambda^2$, we obtain 
\beq
 G \mu \simeq 6.7 \times 10^{-13} \times c_\mu \, \lmk \frac{\Lambda}{10^{13} \GeV} \rmk^2. 
\eeq
See \cite{Athenodorou:2021qvs} for an estimate of $c_\mu$ in lattice simulations.

Owing to the structure of the center symmetry for SO($2N$), the existence of one F-string and two D-strings is confirmed without the presence of a new compsite state. For instance, the collision of two different D-strings leads to the formation of an F-string (see Ref.~\cite{Yamada:2022imq,Yamada:2022aax} for detail). 
This contrasts with cosmic superstrings in brane inflationary scenarios~\cite{Dvali:2003zj,Copeland:2003bj} (see also Refs.~\cite{Polchinski:1988cn,Jackson:2004zg,Hanany:2005bc}), where composite strings consist of multiple F-strings and D-strings. 
Thus, our model exhibits greater simplicity and can be described by a couple of VOS equations (for F- and D-strings) with an interaction term (see Refs.~\cite{Avgoustidis:2007aa,Rajantie:2007hp,Pourtsidou:2010gu} for this formalizm). 
Furthermore, a hierarhy may exist between the intercommutation probabilities of F- and D-strings, which allows further simplification of the cosmic-string network by employing a single type of cosmic string. 
%Furthermore, a hierarhy may exist between the intercommutation probabilities of F- and D-strings. 
A cosmic string with a lower intercommutation probability is prelvalent in the scaling regime and dominates the network. 
In this paper, we therefore focus on the dynamics of 
a single type of cosmic string with a low intercommutation probability.

\subsection{Gravitational waves from cosmic strings}
Typically, the complex dynamics of cosmic strings yield GWs~\cite{Vilenkin:1981bx,Vachaspati:1984gt}. 
We solve the dynamics of cosmic strings using an extended version of velocity-dependent one-scale (VOS) model~\cite{Kibble:1984hp,Martins:1995tg,Martins:1996jp,Martins:2000cs}  
as introduced in Ref.~\cite{Avgoustidis:2005nv} and further explained in Ref.~\cite{Yamada:2022imq}, 
and calculate GW spectrum emitted from cosmic strings~\cite{Caldwell:1991jj,DePies:2007bm,Sanidas:2012ee,Sousa:2013aaa,Sousa:2016ggw}. 
This extended model is based on reasonable assumptions that qualitatively agree with numerical simulations~\cite{Avgoustidis:2005nv}. 
The major difference from the original VOS model is to treat the inter string distance and the correlation length of cosmic strings separately for a small $P$. 
Further, the effective intercommutation probability $P_{\rm eff}$ is effectively enhanced by an order of $10$, because cosmic strings possess a small wiggly structure that leads to multiple intersections within a short time~\cite{Avgoustidis:2005nv}. 
With regard to the size of the string loops, 
we adopt the widely-accepted assumption of $\alpha =0.1$, which is confirmed by the Nambu-Goto simulations with $P =1$~\cite{Blanco-Pillado:2013qja} (see also Refs.~\cite{Ringeval:2005kr,Blanco-Pillado:2011egf,Blanco-Pillado:2013qja,Blanco-Pillado:2017oxo,Blanco-Pillado:2017rnf}). 
See Ref.~\cite{Yamada:2022imq} for the detail of our numerical procedure. 
Moreover, we take into account the dilution of cosmic string loops during the early glueball dominated era, 
which suppresses the amplitude of GW signals at high frequencies.

\begin{figure}
    \centering
            \includegraphics[width=0.95\hsize]{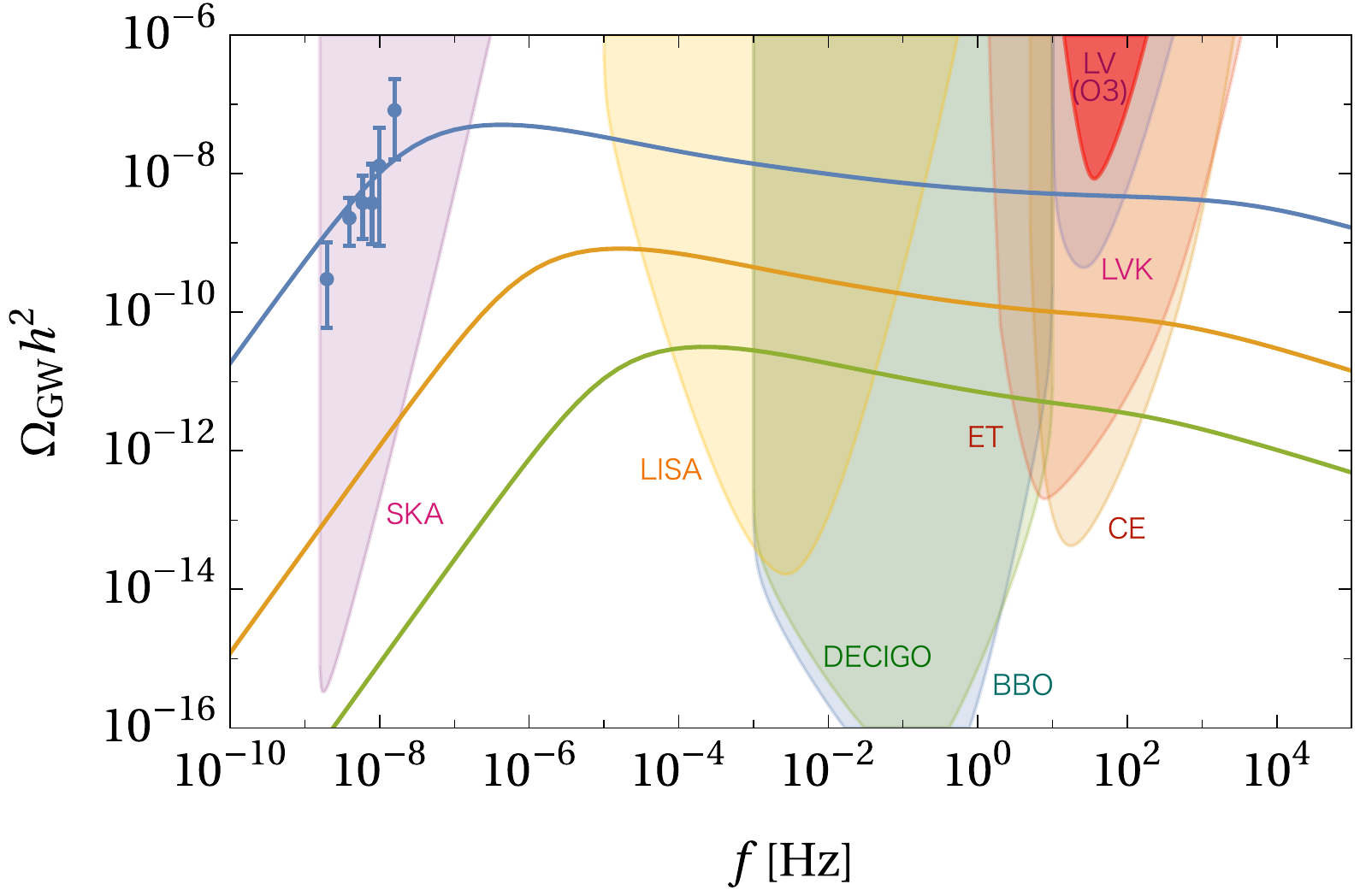}
    \caption{
    GW spectra emitted from cosmic strings in pure SO($2N$) gauge theory for the case with $(G\mu, P) = (4 \times 10^{-12}, 10^{-3})$ (blue), $(10^{-13}, 10^{-2})$ (yellow), $(7 \times 10^{-15}, 10^{-1})$ (green). The region shaded in red is excluded by LV O3 data~\cite{KAGRA:2021kbb,LIGOScientific:2021nrg}. The data points in the upper left corner represent some of the NANOGrav 15-year results~\cite{NANOGrav:2023gor}. The light-shaded regions denote projected sensitivities. 
}
    \label{fig:result}
\end{figure}

Figure~\ref{fig:result} shows three examples of GW spectra originating from cosmic strings for the cases of $(G\mu, P) = (4 \times 10^{-12}, 10^{-3})$ (blue), $(10^{-13}, 10^{-2})$ (yellow), and $(7 \times 10^{-15}, 10^{-1})$ (green). 
The decay temperature of glueballs is assumed to be 
$T_d = 10^5 \GeV$, $10^3 \GeV$, and $10^2 \GeV$, respectively. 
Note that $c_\mu$ can be larger than unity for D-strings by a factor of $\mathcal{O}(N)$, 
so that the green curve can be considered as the lower boundary of our prediction for $\Lambda = 10^{12}\,\text{-}\,10^{14} \GeV$ and $c_\mu \gtrsim 1$. 
We note that the overall amplitude of the GW spectum scales as $P_{\rm eff}^{-1} \sqrt{G\mu}$ and the peak frequency is proportional to $(G\mu)^{-1}$, 
where $P_{\rm eff} = 1 - (1-P)^{N_{\rm scat}}$ with $N_{\rm scat} = 10$ represent the effective intercommutation probabilities for long strings.

The red, densely shaded region is excluded by the advanced LIGO/Virgo's (LV) third observation run (O3) data~\cite{KAGRA:2021kbb,LIGOScientific:2021nrg}. 
The power-law-integrated sensitivity curves for upcoming GW experiments 
are plotted as light shaded regions in accordance with Ref.~\cite{Schmitz:2020syl}. 
These include projects such as 
SKA~\cite{Janssen:2014dka},
LISA~\cite{LISA:2017pwj},
DECIGO~\cite{Kawamura:2011zz,Kawamura:2020pcg},
BBO~\cite{Harry:2006fi},
Einstein Telescope (ET)~\cite{Punturo:2010zz,Maggiore:2019uih},
Cosmic Explorer (CE)~\cite{Reitze:2019iox},
and aLIGO+aVirgo+KAGRA (LVK)~\cite{Somiya:2011np,KAGRA:2020cvd}. 
Even the prediction of the green curve is expected to be observed by some of upcoming GW experiments, including LISA and ET, so that the baryonic glueball DM scenario can be indirectly tested by those experiments.

Recently, observations from 
NANOGrav~\cite{NANOGrav:2023gor}, 
PPTA~\cite{Reardon:2023gzh}, 
EPTA~\cite{Antoniadis:2023ott}, 
and CPTA~\cite{Xu:2023wog} 
have reported evidence supporting the existence of a stochastic GW signal consistent with the Hellings-Downs correlation. 
Select data from NANOGrav's 15-year observations~\cite{NANOGrav:2023gor} are depicted in the upper left corner to demonstrate the favored region. 
For a more streamlined representation, only six frequency bins that appeared to be inconsistent with the null results are included. 
The blue curve is found to be consistent with both the NANOGrav data and the LKV O3 data, suggesting $\Lambda \sim 10^{13}$ and $N \sim 10$. 
This is consistent with the recent study~\cite{Ellis:2023tsl} for cosmic superstrings~\cite{Polchinski:1988cn,Dvali:2003zj,Copeland:2003bj,Jackson:2004zg,Hanany:2005bc}.%
\footnote{
For cosmic string interpretation of PTA, see also Refs.~\cite{Ellis:2020ena,Blasi:2020mfx,Samanta:2020cdk,Blanco-Pillado:2021ygr,NANOGrav:2023hvm,Antoniadis:2023zhi,Kitajima:2023vre,Wang:2023len,Bian:2023dnv,Lazarides:2023ksx,Eichhorn:2023gat,Figueroa:2023zhu,Wu:2023hsa,Antusch:2023zjk,Buchmuller:2023aus}. 
}
This result would be tested by upcoming runs of LVK and other GW detection experiments.

In Fig.~\ref{fig:result}, it can be observed that the GW spectrum at frequencies higher than approximately $10^2 \, {\rm Hz}$ is slightly attenuated by the entropy produced by glueball decay. The attenuation factor is more pronounced for a lower glueball decay rate, which requires a smaller $\Lambda$. 
If we take $\Lambda \sim 10^{13} \GeV$ to account for the NANOGrav 15-year data 
and take $M \lesssim \Mpl$, 
the attenuation factor does not have significant relevance to the frequencies around the constraint from the LV O3 data.

\section{Discussion and conclusions}
\label{sec:discussion}
We have pointed out that a pure SO($2N$) gauge theory predicts the formation of cosmic strings (macrosropic color flux tubes) as well as baryonic glueballs after the confinement phase transition. The baryonic glueballs are stable and are expected to form at the confinement phase transition via a variant of the Kibble-Zurek mechanism. 
Our findings suggest that 
baryonic glueballs can account for the abundance of DM when $\Lambda = \mathcal{O}(10^{12}\,\text{-}\,10^{13}) \GeV$. 
We have calculated the GW spectrum and have found that our model can be tested by future GW observations, including those from LISA and ET, 
even considering $\mathcal{O}(1)$ uncertainties. 
Notably, 
the NANOGrav 15-year data can be consistently explained by the baryonic glueball DM scenario, inferring $N \sim 10$ and $\Lambda = 10^{13}\GeV$. 
One may think that those values appear quite reasonable from the point of view of string theory. 
Note that the renormalization group equation implies $\alpha^{-1} \simeq 4 (N-1)$ around the GUT scale for $\Lambda = 10^{13} \GeV$, which is close to the SM gauge coupling constant for $N \sim 10$. 
Additionaly, moderately large gauge groups often appear in string theory.
For instance, some superstring theories employ a gauge group $\SO(32)$ (or more precisely, $\Spin(32)/\bZ_2$) in ten dimensions.

Our calculations inherently contain several theoretically unknown $\mathcal{O}(1)$ factors, owing to the strong dynamics. 
Some of these $\mathcal{O}(1)$ factors may be determined by quantum lattice simulations for the SO($2N$) gauge theory. 
Additional uncertainties in our prediction originate from the dynamics of cosmic strings with small intercommutation probabilities. 
In principle, the dynamics of the cosmic string network can be determined by a Nambu-Goto simulation; however, it is quite notably challengins to conduct detailed numerical simulations for a small intercommutation probability. 
Further studies employing such comprehensive numerical simulations are required to quantitatively enhance our predictions. 

Despite these challenges, the 
qualitative analysis in this paper remains robust. 
The stability of cosmic strings and baryonic glueballs is guranteed by (generalized) symmetry that recently develops in theoretical physics. 
Their formation at the confinement phase transition is also plausible due to causality. 
Consequently, if the GW signals reported by the NANOGrav 15-year data are indeed derived from cosmic strings, it is reasonalbe to expect that the DM are baryonic glueballs. 
This can be indirectly verified by analyzing the GW spectrum across a broader frequency range in future GW detection experiments, including those by LVK and LISA.

\

\section*{Acknowledgments}
The present work is supported by JSPS KAKENHI Grant Numbers  20H05851 (M.Y.), 23K13092 (M.Y.), 21H05188 (K.Y.), 17K14265 (K.Y.), and JST FOREST Program Grant Number JPMJFR2030 (K.Y.). MY was supported by MEXT Leading Initiative for Excellent Young Researchers. 

%%%%%%%%%%%%%%%%%%%%%%%%%%%%%%%%%%%%%%%%%%%%%%%%%%%%%%%%%%%%%%%%%%%%%%%%%%%%%%%%%%%%%%%%%%%%%%%%%%%%

\bibliographystyle{JHEP}
\bibliography{ref}

%%%%%%%%%%%%%%%%%%%%%%%%%%%%%%%%%%%%%%%%%%%%%%%%%%%%%%%%%%%%%%%%%%%%%%%%%%%%%%%%%%%%%%%%%%%%%%%%%%%%

\end{document}